\pdfoutput=1
\documentclass[a4paper,11pt]{article}

\usepackage[
    margin=0.8in
    ]{geometry}

\usepackage{amsmath,amssymb,amsthm,mathtools,mathrsfs,amsfonts,latexsym}
\usepackage[italicdiff]{physics}
\mathtoolsset{showonlyrefs,showmanualtags}
\numberwithin{equation}{section}
\allowdisplaybreaks[4]

\usepackage[
    bibstyle=phys,
    sorting=none,
    biblabel=brackets,
    citestyle=numeric-comp,
    pageranges=false,
    eprint=true,
    url=true,
    ]{biblatex}
\DeclareFieldFormat{url}{[\href{#1}{iNSPIRE}]}
\DeclareSourcemap{
	\maps[datatype=bibtex]{
		\map[overwrite=true]{
			\step[fieldsource=shortjournal]
			\step[fieldset=journal, origfieldval]
		}
	}
}
\AtEveryBibitem{
    \iffieldequalstr{type}{NoiNSPIRE}{
        \DeclareFieldFormat{url}{[\href{#1}{Link}]}
        }{}
    \clearfield{note}
    \clearfield{type}
    \clearname{editor}
    \clearlist{location}
}
\bibliography{reference}

\usepackage{tikz}
\usetikzlibrary{intersections,calc}
\usepackage[subrefformat=parens]{subcaption}

\usepackage{hyperref}
\hypersetup{
    colorlinks=true,
    linktocpage=true,
    linkcolor=blue,
    urlcolor=blue,
    citecolor=blue,
    filecolor=blue,
    bookmarksopen=true,
    }

\makeatletter
\renewcommand\thefootnote{\@fnsymbol\c@footnote}%
\renewcommand{\maketitle}{%
    \newpage
    \begin{flushright}
        UTHEP- 782
    \end{flushright}
    \null
    \vskip 2em%
    \begin{center}%
        {\LARGE\textbf{\@title}\par}%
        \vskip 1.5em%
        {\large%
            \lineskip .5em%
            \begin{tabular}[t]{c}%
                \@author
            \end{tabular}\par}%
        \vskip 1em%
        {\large \@date}%
    \end{center}%
    \par%
    \@thanks%
    \vskip 1.5em}
\makeatother
\usepackage{authblk}

\title{Closed string amplitudes around tachyon vacuum solution in Kaku theory}
\author{Yuji Ando\thanks{E-mail: \href{mailto:ando@het.ph.tsukuba.ac.jp}{ando@het.ph.tsukuba.ac.jp}}}
\affil{
Degree Programs in Pure and Applied Sciences,\\
Graduate School of Science and Technology, University of Tsukuba,\\
Tsukuba, Ibaraki 305-8571, Japan}
\date{}

\begin{document}
\maketitle
\renewcommand\thefootnote{\arabic{footnote}}
\setcounter{footnote}{0}
\begin{abstract}
    We incorporate closed string field into Kaku's open string field theory which is defined using Kaku vertex, and we construct open-closed string field theory. To do this, we define new consistent open-closed vertex and open-open-closed vertex with the Kaku vertex. Because these vertices depend on Chan-Paton parameter such as the Kaku vertex, the open-closed string field theory action that we construct depends on the Chan-Paton parameter such as the Kaku's theory action. However, we can show that an infinitesimal change in $l$ corresponds to a field redefinition. Furthermore, we compute closed string amplitudes around tachyon vacuum solution in this theory. As a result, we confirm that these amplitudes are conventional pure closed string amplitudes on surfaces without boundaries.
\end{abstract}
\thispagestyle{empty}
\newpage
\setcounter{page}{1}
\tableofcontents
\section{Introduction and Summary}
In string field theory, which is a candidate for a non-perturbative formulation of string theory, every classical solution represents a string background. In particular, one nontrivial solution is tachyon vacuum solution. There are no open string excitations around this solution and only closed strings should exist. Hence, it is expected that closed string amplitudes are closed string amplitudes on surfaces without boundaries, namely, pure closed string amplitudes.

One of bosonic string field theories is Witten theory \cite{Witten:1985cc}. In this theory, tachyon vacuum solutions were constructed \cite{Schnabl:2005gv,Okawa:2006vm,Erler:2009uj,Erler:2010zza,Ellwood:2006ba}. However, we cannot compute closed string amplitudes. This is because the theory gives a single cover of moduli space even though the Witten theory action is given only by open string field and its interaction \cite{Zwiebach:1990az}. Hence it is unknown how to compute closed string amplitudes without modification. As a result, even though the tachyon vacuum solution has been found, it is unclear how closed strings appear around the tachyon vacuum solution. If we modify the Witten vertex to something else, or if other modifications were made, it might be possible to compute closed string amplitudes, but currently the tachyon vacuum solution has only been found in the Witten theory.

Another theory is the theory that action is given by light-cone vertex like HIKKO theory or Kugo-Zweibach theory \cite{Hata:1986jd,Hata:1986ke,Kugo:1992md,Saitoh:1988ew,Kugo:1997rm,Asakawa:1998em}. Unlike the Witten theory, in this theory, the action is given by both the open string field and the closed string field, making it easier to compute closed string amplitudes. However the tachyon vacuum solution has not yet been found, and we cannot compute closed string amplitudes around it.

Our purpose is to find a tachyon vacuum solution and to compute closed string amplitudes around the solution. Since we face problems in the above theories, we have to look for a theory where we do not face any problems. As a candidate for this purpose, we focus on the theory  proposed by Kaku \cite{Kaku:1987jx}\footnote{Recently, string field theory using hyperbolic metric has been actively studied \cite{Moosavian:2017qsp,Moosavian:2017sev,Costello:2019fuh,Cho:2019anu,Firat:2021ukc,Ishibashi:2022qcz,Firat:2023suh}. In this paper, we do not focus such a construction but \cite{Bernardes:2024ncs} builds upon the developments in the hyperbolic vertex to construct the Kaku vertex.} and we refer to it as open Kaku theory in this paper. The open Kaku theory action is given by open string field and open string Kaku vertex. The open string Kaku vertex is defined using a parameter $l$, called the Chan-Paton parameter. For specific choices of $l\to\infty$ and $l\to0$, the open Kaku theory action coincides with the Witten theory action and the action given by the light-cone vertex respectively. Thus, the open Kaku theory is intermediate between these two, and this theory may be what we are looking for. In fact, tachyon vacuum solutions have been already found \cite{Ando:2023cnx}. Currently, the open Kaku theory describes open string physics but does not describe closed string physics. Therefore, the remaining task is to incorporate closed string field into the open Kaku theory and to construct open-closed string field theory. In this theory, we can compute closed string amplitudes around the tachyon vacuum solution.

In this paper, we complete the task of incorporating the closed string field and constructing a new open-closed string field theory, which we call open-closed Kaku theory. Because an open-closed string field theory should have open-closed homotopy algebra (OCHA) structure \cite{Zwiebach:1997fe,Kajiura:2004xu,Kajiura:2005sn}, we require that the open-closed Kaku theory satisfies this condition. As a result, we obtain the action for the open-closed Kaku theory. The theory includes open-closed Kaku vertex, open-open-closed Kaku vertex and closed string light-cone vertex as closed string interaction vertex. The open-closed Kaku vertex and the open-open-closed Kaku vertex depends on the Chan-Paton parameter, and for the specific choices of $l\to0,\infty$, these vertices coincide with the interaction vertex proposed in \cite{Saitoh:1988ew,Kugo:1997rm,Asakawa:1998em} and the interaction vertex proposed by \cite{Shapiro:1987ac,Zwiebach:1992bw}, respectively. Furthermore, following \cite{Erler:2020beb}, we show that an infinitesimal change of $l$ corresponds to a field redefinition.

Finally, we compute the closed string tree amplitudes around the tachyon vacuum solution and confirm that they are pure closed string amplitudes. Usually, when computing amplitudes, it is necessary to choose a gauge condition and a propagator. Surprisingly, by using only one of the OCHA relations and the fact that there are no open string excitations around the tachyon vacuum solution, we obtain that closed string amplitudes around the tachyon vacuum solution are pure closed string amplitudes. This derivation does not depend on the choice of gauge condition.

This paper is organized as follows. In section \ref{def:vertex}, we define the open-closed Kaku vertex and the open-open-closed Kaku vertex and construct the open-closed Kaku theory. We also show that the open-closed Kaku theory is independent of $l$. In section \ref{amplitude}, we confirm that the closed string amplitudes around the tachyon vacuum solution are as expected. Finally, in section \ref{Summary}, we present the summary.
\section{Open-closed Kaku theory\label{def:vertex}}
Let us consider an open string field $\Psi$ and a closed string field $\Phi$ where the closed string field satisfies the level-matching condition. We consider the following action
\begin{align}
    S=\sum_{p,q\ge0}\frac{1}{p!(q+1)}\omega_o(\Psi,n_{p,q}(\Phi^{\wedge p};\Psi^{\otimes q}))+\sum_{k\ge1}\frac{1}{(k+1)!}\omega_c(\Phi,L_k(\Phi^{\wedge k})),
\end{align}
where $\omega_o$ and $\omega_c$ are odd symplectic forms. They are defined by BPZ inner product as follows\footnote{See \cite{Kunitomo:2022qqp} for more detail relation between them and BPZ inner product.},
\begin{align}
    \omega_o(\Psi_1,\Psi_2)=(-1)^\abs{\Psi_1}\ev{\Psi_1,\Psi_2}\qc\omega_c(\Phi_1,\Phi_2)=(-1)^\abs{\Phi_1}\ev{\Phi_1,c_0^-\Phi_2},
\end{align}
where $\abs{\Psi}$ is defined as its Grassmann parity plus 1 and $\abs{\Phi}$ is defined as its Grassmann parity\footnote{We follow the notation in \cite{Erler:2015uba}.}. $n_{p,q}$ and $L_k$ are multilinear map defined on the first-quantized state space of open string $\mathcal{H}_o$ and the first-quantized state space of closed string $\mathcal{H}_c$, respectively.
\begin{align}
    L_k&:\mathcal{H}_c^{\otimes k}\to\mathcal{H}_c\\
    n_{p,q}&:\mathcal{H}_c^{\wedge p}\otimes\mathcal{H}_o^{\otimes q}\to\mathcal{H}_o
\end{align}
In particular, $n_{0,1}$ and $L_1$ are the BRS operators $Q_o,Q_c$ for open strings and closed strings, respectively. To satisfy Batalin-Vilkovisky master equation, open-closed string field theory must have cyclic open-closed homotopy structure\footnote{In \cite{Maccaferri:2022yzy}, Sphere-Disk homotopy algebra (SDHA) is introduced, and it is in turn an extension of the OCHA. In this paper, we consider only the OCHA structure and do not consider the SDHA structure.}. This implies that the multilinear maps $n_{p,q},L_k$ and the odd symplectic forms $\omega_o,\omega_c$ must satisfy four equations \cite{Kajiura:2004xu,Kajiura:2005sn}. Two of them are $L_\infty$ relation and OCHA relation\footnote{In this paper, we set $n_{0,0}$ and $L_0$ to zero maps and do not consider weak OCHA.}
\begin{align}
    0&=\sum_\sigma\sum_{k=1}^m(-1)^{\epsilon(\sigma)}\frac{1}{k!(m-k)!}L_{m-k+1}(L_k\qty(\Phi_{\sigma(1)},\dots,\Phi_{\sigma(k)}),\Phi_{\sigma(k+1)},\dots,\Phi_{\sigma(m)}),\label{Linfty}\\
    0&=\sum_\sigma\sum_{k=1}^m(-1)^{\epsilon(\sigma)}\frac{1}{k!(m-k)!}n_{m-k+1,l}\qty(L_k\qty(\Phi_{\sigma(1)},\dots,\Phi_{\sigma(k)}),\Phi_{\sigma(k+1)},\dots,\Phi_{\sigma(m)};\Psi_1,\dots,\Psi_l)\\
    &\qquad+\sum_\sigma\sum_{k=0}^m\sum_{j=0}^l\sum_{i=0}^{l-j}(-1)^{\mu_{k,i}(\sigma)}\frac{1}{k!(m-k)!}\\
    &\qquad\quad\times n_{k,l-j+1}\qty(\Phi_{\sigma(1)},\dots,\Phi_{\sigma(k)};\Psi_1,\dots,\Psi_i,n_{m-k,j}\qty(\Phi_{\sigma(k+1)},\dots,\Phi_{\sigma(m)};\Psi_{i+1},\dots,\Psi_{i+j}),\Psi_{i+j+1},\dots,\Psi_l).\label{OCHA}
\end{align}
Here $\mu_{m,i}$ is given by
\begin{align}
    \mu_{k,i}(\sigma)\coloneqq\epsilon(\sigma)+\sum_{j=1}^k\abs{\Phi_{\sigma(j)}}+\sum_{j=1}^i\abs{\Psi_j}\qty(1+\sum_{n=k+1}^m\abs{\Phi_{\sigma(n)}}),
\end{align}
and $\epsilon(\sigma)$ is Koszul sign defined by
\begin{align}
    \Phi_1\wedge\dots\wedge\Phi_n=(-1)^{\epsilon(\sigma)}\Phi_{\sigma(1)}\wedge\dots\wedge\Phi_{(\sigma(n)}.
\end{align}
The remaining conditions are the cyclic conditions
\begin{align}
    \omega_c(\Phi_1,L_k(\Phi_2,\dots,\Phi_{k+1})&=-(-1)^{\abs{\Phi_1}}\omega_c(L_k(\Phi_1,\dots,\Phi_k),\Phi_{k+1}),\\
    \omega_o(\Psi_1,n_{p,q}(\Phi_1,\dots,\Phi_p;\Psi_2,\dots,\Psi_{q+1}))&=-(-1)^{\abs{\Psi_1}(\abs{\Phi_1}+\dots+\abs{\Phi_p})}\omega_o(n_{p,q}(\Phi_1,\dots,\Phi_p;\Psi_1,\dots,\Psi_q),\Psi_{q+1}).\label{cOCHA}
\end{align}

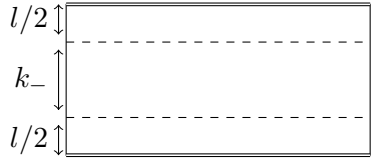
\begin{figure}
    \centering
    \begin{tikzpicture}
        \draw (-2,0)--(2,0)--(2,2)--(-2,2)--cycle;
        \draw[dashed] (-2,1/2)--(2,1/2);
        \draw[dashed] (-2,3/2)--(2,3/2);
        \draw[<->,shift={(-0.1,0)}] (-2,3/2+0.1)--(-2,2) node[midway,left] {$l/2$};
        \draw[<->,shift={(-0.1,0)}] (-2,1/2+0.1)--(-2,3/2-0.1) node[midway,left] {$k_-$};
        \draw[<->,shift={(-0.1,0)}] (-2,0)--(-2,1/2-0.1) node[midway,left] {$l/2$};
        \draw[double] (-2,0)--(2,0);
        \draw[double] (2,2)--(-2,2);
    \end{tikzpicture}
    \caption{An open string propagator with the length $k_-+l$}\label{Kakustring}
\end{figure}

In this section, we incorporate the closed string field into the open Kaku theory. The open Kaku theory action is given by open string Kaku vertex. To define this vertex, we separate the length of the open string propagator into two parts: the momentum $k_-$ part and the Chan-Paton parameter $l$ part (Figure \ref{Kakustring}).
The Kaku vertex is defined by gluing in the same way as in the Witten vertex for the Chan–Paton parameter $l$ part, and in a way analogous to the light-cone vertex for the momentum $k_-$ part \cite{Kaku:1987jx}. See \cite{Erler:2020beb,Ando:2023cnx} for more details. Using the cubic and quartic Kaku vertices $n_{0,2}^l,n_{0,3}^l$, the Kaku theory action is given by
\begin{align}
    S=\frac{1}{2}\omega_o(\Psi,Q_o\Psi)+\frac{1}{3}\omega_o(\Psi,n_{0,2}^l(\Psi,\Psi))+\frac{1}{4}\omega_o(\Psi,n_{0,3}^l(\Psi,\Psi,\Psi)).
\end{align}
Our first task is to find appropriate vertices $n_{p,q}^l$ for $p\ge1,q\ge0$ and $L_k^l$ for $k\ge2$ such that they satisfy the four relations.
\subsection{Open-closed Kaku vertex\label{open-closed}}
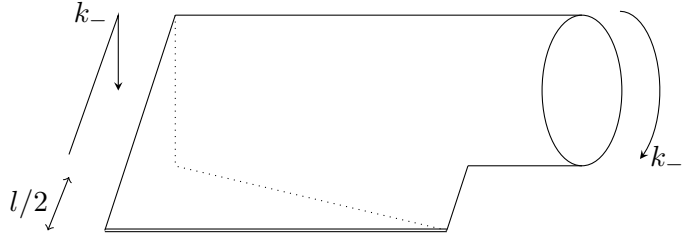
\begin{figure}
    \centering
    \begin{tikzpicture}[scale=3]
        \coordinate (o) at (0,0);
        \coordinate (i) at ($(o)+(pi*2/5 r:0.3)+(1.5,0)$);
        \coordinate (i2) at ($(i)+(0,{0.7*sin(pi*2/5 r)})$);
        \coordinate (v) at ($(i)+(-pi*3/5 r:0.3)$);
        \coordinate (s1) at ($(o)+(0,0)$);
        \coordinate (p11) at ($(i)+(-1.5,0)$);
        \coordinate (p1c) at ($(p11)+(pi*2/5 r:0.7)$);
        \coordinate (p12) at ($(p1c)+(-pi/2 r:1/3)$);
        \coordinate (s2) at ($(p12)+(-pi/2 r:1/3)$);
        \draw[<->] ($(s1)+(-1/4,0)$)--($(i)+(-1.5,0)+(-1/4,0-0.05)$) node[midway,left] {$l/2$};
        \draw[->,>=stealth] ($(i)+(-1.5,0)+(-1/4,0+0.05)$)--($(p1c)+(-1/4,0)$) node[left] {$k_-$}--($(p12)+(-1/4,0)$);
        \draw ($(i)!1/2!(i2)+(1/2,0)$) ellipse (1/2*0.7/2 and {0.7*sin(pi*2/5 r)/2});
        \draw[->,>=stealth,domain=pi/2:-pi/3,shift={($(i)!1/2!(i2)+(2/3,0)$)}] plot({1/2*0.7/2*cos(\x r)},{0.7/2*sin(\x r)}) node[right]{$k_-$};
        \draw[double] (s1)--(v);
        \draw[dotted] (v)--(s2);
        \draw (s1)--(p1c);
        \draw[dotted] (p1c)--(s2);
        \draw (p1c)--(i2)--++(1/2,0);
        \draw (i)--++(1/2,0);
        \draw (v)--(i);
    \end{tikzpicture}
    \caption{The open-closed Kaku vertex}\label{open-closedKaku}
\end{figure}

First, we construct the open-closed Kaku vertex. The open-closed Kaku vertex should coincide with the light-cone open-closed vertex \cite{Saitoh:1988ew,Asakawa:1998em} in the limit $l\to0$, whereas it should coincide to the open-closed vertex \cite{Shapiro:1987ac,Zwiebach:1992bw} in the limit $l\to\infty$. To define the Kaku vertex, we assign only the momentum $k_-$ as the length of the closed string propagator. Then the open-closed Kaku vertex is represented by Figure \ref{open-closedKaku}.
The surface in the figure satisfies the requirement in the limit $l\to0,\infty$ respectively.

Let us write the vertex in term of a correlation function. Given an open string state $\ket{A}$, we can define a corresponding boundary operator $A(0)$ such that
\begin{align}
    \ket{A}=A(0)\ket{0}
\end{align}
where $\ket{0}$ is the $SL(2,\mathbb{R})$ invariant vacuum. Similarly given a closed string state $\ket{A_c}$, we can define a corresponding bulk operator $A_c(0)$. Then using the boundary operator and the bulk operator, we can write the vertex as
\begin{align}
    V_{(1,1)}^l(B_c,A)&=\ev{f_{((1,1),c)}^l\circ B_c(0)f_{((1,1),o)}^l\circ A(0)}_\mathrm{UHP}\\
    &=(-1)^{1+(\abs{A}+1)(\abs{B_c}+1)}\omega_o(A,n_{1,0}^l(B_c)).
\end{align}
The maps $f_{((1,1),o)}^l,f_{((1,1),c)}^l$ are found from a mapping from the complex coordinate $u$ on upper half plane (UHP) to the complex coordinate on the surface in Figure \ref{open-closedKaku}. Suppose that the open string state $A$ has a momentum $k_-$. As a result of momentum conservation, the closed string state $B_c$ has the momentum $-k_-$. When we write a complex coordinate on the surface in Figure \ref{open-closedKaku} which is normalized by $\abs{k_-}$ as $\rho$, the mapping satisfies\footnote{See Section 5 and Appendix B in \cite{Erler:2020beb} for how to construct local coordinate map.}
\begin{align}
    \rho=\frac{1}{\pi}\int\dd{u}\frac{1}{\abs{u_*}}\frac{\sqrt{(u-u_*)(u-\bar{u}_*)}}{u(u-i)(u+i)},
\end{align}
where the operators $A$ and $B_c$ are placed respectively $0$ and $i$ on UHP. $u_*$ is the position of the interaction point, and it is restricted to pure imaginary since the length parameter of the closed string should be a real parameter. Hence we can write
\begin{align}
    u_*=iy\qc y\in\mathbb{R}.
\end{align}
Then $\rho$ is related to $u$ through
\begin{align}
    \rho&=\frac{1}{\pi}\int\dd{u}\frac{1}{y}\frac{\sqrt{(u-iy)(u+iy)}}{u(u-i)(u+i)}\\
    &=\frac{1}{\pi}\ln(\frac{u}{i\sqrt{u^2+y^2}+iy})-\frac{\sqrt{y^2-1}}{2\pi y}\ln(\frac{-u^2-1}{\qty(\sqrt{y^2-1}+\sqrt{u^2+y^2})^2}),
\end{align}
where we choose an integration constant such that $\rho(u_*)=0$ holds. The momentum $\abs{k_-}$ and the Chan-Paton parameter $l$ are related to the position of the interaction point $iy$ by
\begin{align}
    \frac{\abs{k_-}}{\abs{k_-}+l}=\frac{\sqrt{y^2-1}}{y}.
\end{align}
Solving the equation gives
\begin{align}
    y=\frac{1+L}{\sqrt{L(2+L)}},\label{relation:yl}
\end{align}
where
\begin{align}
    L\coloneqq\frac{l}{\abs{k_-}}.
\end{align}

By appropriate scaling, translation, and exponentiation of $\rho$, we can find the inverse local coordinate maps given by
\begin{align}
    (f_{((1,1),o)}^l)^{-1}&=\frac{u}{\sqrt{u^2+y^2}+y}\qty(\frac{u^2+1}{\qty(\sqrt{y^2-1}+\sqrt{u^2+y^2})^2})^{-\sqrt{y^2-1}/2y},\\
    (f_{((1,1),c)}^l)^{-1}&=\frac{-u^2-1}{\qty(\sqrt{y^2-1}+\sqrt{u^2+y^2})^2}\qty(\frac{u}{i\sqrt{u^2+y^2}+iy})^{-2y/\sqrt{y^2-1}}
\end{align}
and their derivatives are
\begin{align}
    \eval{\dv{(f_{((1,1),o)}^l)^{-1}}{u}}_{u=0}&=\frac{(y+\sqrt{y^2-1})^{{\sqrt{y^2-1}/y}}}{2y},\\
    \eval{\dv{(f_{((1,1),c)}^l)^{-1}}{u}}_{u=i}&=\frac{1}{2i(y^2-1)}\qty(\frac{1}{\sqrt{y^2-1}+y})^{-2y/\sqrt{y^2-1}}.
\end{align}

Let us consider the limit $y\to1$. Because of \eqref{relation:yl}, the limit $y\to1$ corresponds to the limit $l\to\infty$ or $\abs{k_-}\to0$. Then the derivatives of the inverse local coordinate maps become
\begin{align}
    \lim_{y\to1}\eval{\dv{(f_{((1,1),o)}^l)^{-1}}{u}}_{u=0}&=\frac{1}{2},\label{oc:w}\\
    \lim_{y\to1}\eval{\dv{(f_{((1,1),c)}^l)^{-1}}{u}}_{u=i}&=-i\infty.
\end{align}
Hence the derivative of $(f_{((1,1),o)}^l)^{-1}$ agrees with the result obtained in \cite{Ellwood:2008jh} for the same quantity\footnote{More concretely, \eqref{oc:w} matches with $(dw/dz)^{-1}$ where $w(z)$ is given equation (2.1) in \cite{Ellwood:2008jh}. }. On the other hand, the derivative of $(f_{((1,1),c)}^l)^{-1}$ is not well-defined. This is not surprising, as we have already encountered a similar issue with the open string cubic Kaku vertex. Since the cubic light-cone vertex is not well-defined for the state with $k_-=0$, the cubic Kaku vertex in the limit $l\to0$ is also not well-defined. Additionally, for the open-closed Kaku vertex, the open-closed Kaku vertex with any Chan-Paton parameter $l$ is not well-defined, and we need to restrict the closed string state with $k_-=0$ to an on-shell state. Furthermore, in the limit $l\to\infty$,  any closed string state must also be restricted to an on-shell state.

We have defined the open-closed Kaku vertex. Let us now construct an open-closed string field theory action using the open-closed Kaku vertex. We have assigned only the momentum $k_-$ as the length of the closed string propagator. Hence, we consider a theory whose action is given by
\begin{align}
    S&=\frac{1}{2}\omega_o(\Psi,Q_o(\Psi))+\frac{1}{3}\omega_o(\Psi,n^l_{0,2}(\Psi,\Psi))+\frac{1}{4}\omega_o(\Psi,n^l_{0,3}(\Psi,\Psi,\Psi))\\
    &\qquad+\omega_o(\Psi,n^l_{1,0}(\Phi))\\
    &\qquad+\frac{1}{2}\omega_c(\Phi,Q_c(\Phi))+\frac{1}{3!}\omega_c(\Phi,L^\mathrm{lc}_2(\Phi,\Phi)).
\end{align}
where $L^\mathrm{lc}_2$ is closed string light-cone vertex. However, we immediately observe that the theory does not have the OCHA structure. Specifically\footnote{Note that it is sufficient to consider the OCHA relation for $m\ge1$ because the OCHA relation by setting $m=0$ is $A_\infty$ relation and the open Kaku theory has the $A_\infty$ structure.}, by setting $m=1$ and $l=1$, \eqref{OCHA} becomes
\begin{align}
    0&=n_{0,1}\qty(n_{1,1}(A_c;B))+n_{1,1}\qty(L_1(A_c);B)+(-1)^{\abs{A_c}}n_{1,1}\qty(A_c;n_{0,1}(B))\\
    &\qquad+n_{0,2}\qty(n_{1,0}(A_c),B)+(-1)^{\abs{B}}n_{0,2}\qty(B,n_{1,0}(A_c)).\label{OCHA1}
\end{align}
If we set $n_{0,1}(L_1)$ to the BRS operator $Q_o(Q_c)$ for open (closed) string field, $n_{0,2}$ and $n_{1,0}$ to the Kaku vertices $n^l_{0,2},n^l_{1,0}$ and $n_{1,1}$ to zero map, \eqref{OCHA1} becomes
\begin{align}
    0=n_{0,2}^l\qty(n_{1,0}^l(A_c),B)+(-1)^{\abs{B}}n_{0,2}^l\qty(B,n_{1,0}^l(A_c)).
\end{align}
Since this equation does not hold, we cannot set $n_{1,1}$ to the zero map and must construct an open-open closed Kaku vertex that satisfies \eqref{OCHA1}.

\subsection{Open-open-closed Kaku vertex\label{open-open-closed}}
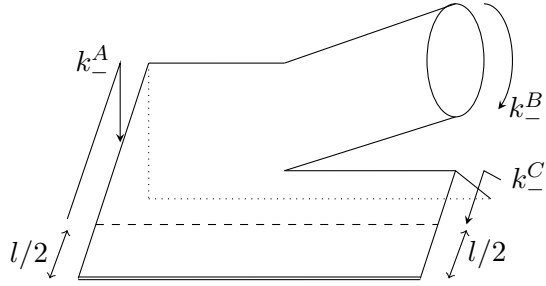
\begin{figure}
    \centering
    \begin{tikzpicture}[scale=1.5]
        \coordinate (o) at (0,0);
        \coordinate (i) at ($(o)+(pi*2/5 r:1)+(1.5,0)$);
        \coordinate (i2) at ($(i)+(0,{sin(pi*2/5 r)})$);
        \coordinate (v) at ($(i)+(-pi*3/5 r:1)$);
        \coordinate (s1) at ($(o)+(0,0)$);
        \coordinate (p11) at ($(s1)+(pi*2/5 r:0.5)$);
        \coordinate (p1c) at ($(p11)+(pi*2/5 r:1.5)$);
        \coordinate (p12) at ($(p1c)+(-pi/2 r:0.7)$);
        \coordinate (s2) at ($(p12)+(-pi/2 r:0.5)$);
        \draw[<->] ($(s1)+(-1/4,0)$)--($(p11)+(-1/4,0-0.05)$) node[midway,left] {$l/2$};
        \draw[->,>=stealth] ($(p11)+(-1/4,0+0.05)$)--($(p1c)+(-1/4,0)$) node[left] {$k_-^A$}--($(p12)+(-1/4,0)$);
        \coordinate (t1) at ($(s2)+(3,0)$);
        \coordinate (p21) at ($(t1)+(pi*4/5 r:0.2)$);
        \coordinate (p2c) at ($(i)+(1.5,0)$);
        \coordinate (p22) at ($(p2c)+(-pi*3/5 r:0.5)$);
        \coordinate (t2) at ($(s1)-(s2)+(t1)$);
        \draw[->,>=stealth] ($(p21)+(1/4,0+0.05)$) node[right] {$k_-^C$}--($(p2c)+(1/4,0)$)--($(p22)+(1/4,0)$);
        \draw[<->] ($(p22)+(1/4,0-0.05)$)--($(t2)+(1/4,0)$) node[midway,right] {$l/2$};
        \draw ($(i)!1/2!(i2)+(1.5,1/2)$) ellipse (1/4 and 1/2);
        \draw[->,>=stealth,domain=pi/2:-pi/3,shift={($(i)!1/2!(i2)+(1.75,1/2)$)}] plot({1/4*cos(\x r)},{1/2*sin(\x r)}) node[right]{$k_-^B$};
        \draw[double] (s1)--(t2);
        \draw[dotted] (s2)--(t1);
        \draw (s1)--(p1c);
        \draw[dotted] (p1c)--(s2);
        \draw (t1)--(p2c)--(t2);
        \draw (p1c)--(i2);
        \draw (i)--(p2c);
        \draw (i2)--($(i)!1/2!(i2)+(1.5,1/2+1/2)$);
        \draw (i)--($(i)!1/2!(i2)+(1.5,1/2-1/2)$);
        %
        \draw[dashed] (p11)--(p22);
    \end{tikzpicture}
    \caption{The open-open-closed Kaku vertex}\label{open-open-closedKaku}
\end{figure}

We define the open-open-closed Kaku vertex, which satisfies \eqref{OCHA1}. The open-open-closed Kaku vertex is represented by Figure \ref{open-open-closedKaku}.
Also, in this case, we will write the vertex in terms of a correlation function. For a point in the moduli space of the Riemann surface, we consider the correlation function
\begin{align}
    \Sigma^{l,m}_{1,2}(B_c,A,C)=\ev{f_{((1,2),c)}^{l,m}\circ B_c(0)f_{((1,2),1)}^{l,m}\circ A(0)f_{((1,2),2)}^{l,m}\circ C(0)}_\mathrm{UHP},
\end{align}
where $m$ is a moduli parameter. These maps $f_{((1,2),r)}^{l,m}$ can be found in a similar way as in the previous section. Using these maps, the open-open-closed Kaku vertex is given by
\begin{align}
    V_{(1,2)}^l(B_c;A,C)&=(-1)^{1+\abs{B_c}+(\abs{A}+1)\abs{B}}\omega_o(A,n_{1,1}^l(B_c;C))\\
    &=\int_{m^l_-}^{m^l_+}\dd{m}\left[\ev{f_{((1,2),1)}^{l,m}\circ b^{l,m}_{((1,2),1),m}f_{((1,2),1)}^{l,m}\circ A(0)f_{((1,2),c)}^{l,m}\circ B_c(0)f_{((1,2),2)}^{l,m}\circ C(0)}_\mathrm{UHP}\right.\\
    &\qquad\qquad\qquad+\ev{f_{((1,2),1)}^{l,m}\circ A(0)f_{((1,2),1)}^{l,m}\circ b^{l,m}_{((1,2),c),m}f_{((1,2),c)}^{l,m}\circ B_c(0)f_{((1,2),2)}^{l,m}\circ C(0)}_\mathrm{UHP}\\
    &\qquad\qquad\qquad\left.+\ev{f_{((1,2),1)}^{l,m}\circ A(0)f_{((1,2),c)}^{l,m}\circ B_c(0)f_{((1,2),1)}^{l,m}\circ b^{l,m}_{((1,2),2),m}f_{((1,2),2)}^{l,m}\circ C(0)}_\mathrm{UHP}\right],
\end{align}
where $b^{l,m}_{((1,2),r),m}$ is the appropriate $b$-ghost insertion for the measure on the moduli space. The region of integration $[m^l_-,m^l_+]$ is determined by the position of the interaction point.

First, we will derive the maps $f_{((1,2),r)}^{l,m}$. Suppose that the states $A,B_c$ and $C$ have momenta $k_-^A,k_-^B$ and $k_-^C$, respectively. In particular, we consider the case of $k_-^A>0,k_-^B,k_-^C<0$\footnote{Of course, it is necessary to define for other cases but we can apply the following discussion to other cases by changing the sign of $\abs{k_-}+l$ or $\abs{k_-}$.}. We write a complex coordinate on the surface in Figure \ref{open-open-closedKaku} as $\rho$. The mapping from the UHP $u$ to the coordinate $\rho$ is given by Mandelstam mapping
\begin{align}
    \rho=\frac{\abs{k_-^A}+l}{\pi}\ln(u)-\frac{\abs{k_-^B}}{2\pi}\ln(u^2+1)-\frac{\abs{k_-^C}+l}{\pi}\ln(u-m),
\end{align}
where the operators $A$ and $B_c$ are placed at $0,i$, respectively. Additionally, we treat the position $m$ of the operator $C$ as the moduli parameter. Since the length parameter of the closed string should be a real parameter, the interaction point $u_*$ satisfies
\begin{align}
    \abs{u_*}^2-2m\Re u_*-1=0.
\end{align}
By appropriate scaling, translation, and exponentiation of $\rho$, we can find the inverse local coordinate maps
given by
\begin{align}
    (f_{((1,2),1)}^{l,m})^{-1}(u)&=\frac{u}{\sqrt{\lambda_1m}}\qty(\frac{\lambda_2\sqrt{m^2+1}}{u^2+1})^{\frac{\lambda_2}{\lambda_1}}\qty(\frac{\sqrt{\lambda_3m(m^2+1)}}{u-m})^{\frac{\lambda_3}{\lambda_1}},\\
    (f_{((1,2),c)}^{l,m})^{-1}(u)&=\qty(\frac{\sqrt{\lambda_1m}}{u})^{\frac{\lambda_1}{\lambda_2}}\frac{u^2+1}{\lambda_2\sqrt{m^2+1}}\qty(\frac{u-m}{\sqrt{\lambda_3m(m^2+1)}})^{\frac{\lambda_3}{\lambda_2}},\\
    (f_{((1,2),2)}^{l,m})^{-1}(u)&=\qty(\frac{\sqrt{\lambda_1m}}{u})^{\frac{\lambda_1}{\lambda_3}}\qty(\frac{u^2+1}{\lambda_2\sqrt{m^2+1}})^{\frac{\lambda_2}{\lambda_3}}\frac{u-m}{\sqrt{\lambda_3m(m^2+1)}},
\end{align}
and their derivatives are
\begin{align}
    \eval{\dv{(f_{((1,2),1)}^{l,m})^{-1}}{u}}_{u=0}&=\frac{1}{\sqrt{\lambda_1m}}\qty(\frac{\sqrt{\lambda_2m(m^2+1)}}{-m})^{\frac{\lambda_2}{\lambda_1}}\qty(\lambda_3\sqrt{m^2+1})^{\frac{\lambda_3}{\lambda_1}},\\
    \eval{\dv{(f_{((1,2),c)}^{l,m})^{-1}}{u}}_{u=i}&=\qty(\frac{\sqrt{\lambda_1m}}{i})^{\frac{\lambda_1}{\lambda_2}}\frac{2i}{\lambda_2\sqrt{m^2+1}}\qty(\frac{i-m}{\sqrt{\lambda_3m(m^2+1)}})^{\frac{\lambda_3}{\lambda_2}},\\
    \eval{\dv{(f_{((1,2),2)}^{l,m})^{-1}}{u}}_{u=m}&=\qty(\frac{\sqrt{\lambda_1m}}{m})^{\frac{\lambda_1}{\lambda_3}}\qty(\frac{m^2+1}{\lambda_2\sqrt{m^2+1}})^{\frac{\lambda_2}{\lambda_3}}\frac{1}{\sqrt{\lambda_3m(m^2+1)}},
\end{align}
where $\lambda_r$ are defined by
\begin{gather}
    \lambda_1=\frac{\abs{u_*}^2}{m}\qc\lambda_2=\frac{2\Re u_*+m(\abs{u_*}^2-1)}{2(1+m^2)}\qc\lambda_3=\frac{\abs{m-u_*}^2}{m(m^2+1)}.
\end{gather}

Next, we will derive $b^{l,m}_{((1,2),r),m}$. This is given by
\begin{align}
    f^{l,m}_{((1,2),r)}\circ b^{l,m}_{((1,2),r),m}=\oint_{f^{l,m}_{((1,2),r)}(0)}\frac{\dd{u}}{2\pi i}v_{((1,2),r),m)}^{l,m}(u)b(u)
\end{align}
where $v_{((1,2),r),m)}^{l,m}(u)$ is defined by
\begin{align}
    v_{((1,2),r),m}^{l,m}(u)&=\pdv{(f_{((1,2),r)}^{l,m})^{-1}}{m}\qty(\pdv{(f_{((1,2),r)}^{l,m})^{-1}}{u})^{-1}\\
    &=\frac{u(u^2+1)}{m(m^2+1)}\qty(\frac{(u_*-m)\Im u_*}{(u-u_*)(u-\bar{u}_*)}+\frac{\Re u_*-m}{u-\bar{u}_*}).
\end{align}
See \cite{Erler:2019loq,Erbin:2021smf} for more details about the derivation. Thus, we obtain
\begin{align}
    V_{(1,2)}^l(B_c;A,C)&=\int_{m^l_-}^{m^l_+}\dd{m}\left\langle\qty(\frac{u_*(u_*^2+1)}{m(m^2+1)}\frac{(u_*-m)}{2}b(u_*)+\frac{\bar{u}_*(\bar{u}_*^2+1)}{m(m^2+1)}\frac{(\bar{u}_*-m)}{2}b(\bar{u}_*))\right.\\
    &\qquad\qquad\qquad\times\left.f_{((1,2),c)}^{l,m}\circ B_c(0)f_{((1,2),1)}^{l,m}\circ A(0)f_{((1,2),2)}^{l,m}\circ C(0)\right\rangle_\mathrm{UHP}.
\end{align}
For the Kaku vertex, the interaction point should be placed in the light-cone portion. This leads to the inequality
\begin{align}
    \frac{l}{2}<\Im\rho(u_*)<\abs{k_-^C}+\frac{l}{2},
\end{align}
and this determines $[m^l_-,m^l_+]$. Particularly, in the limit $l\to\infty$, the inequality is not satisfied, no matter where on the surface $m$ is placed. This is consistent with the action without open-open-closed interaction in \cite{Zwiebach:1992bw}.

Again, we examine whether the OCHA relations hold. Since we construct the open-open-closed vertex such that it satisfies \eqref{OCHA1}, here we consider other cases. First, by setting $m=2,l=0$, \eqref{OCHA} becomes
\begin{align}
    0&=\sum_\sigma(-1)^{\epsilon(\sigma)}\qty(\frac{1}{2}n_{0,1}\qty(n_{2,0}\qty(\Phi_{\sigma(1)},\Phi_{\sigma(2)}))+n_{2,0}\qty(L_1\qty(\Phi_{\sigma(1)}),\Phi_{\sigma(2)}))\\
    &\qquad+\sum_\sigma(-1)^{\epsilon(\sigma)}\qty(\frac{1}{2}n_{1,0}\qty(L_2\qty(\Phi_{\sigma(1)},\Phi_{\sigma(2)}))+(-1)^\abs{\Phi_{\sigma(1)}}n_{1,1}\qty(\Phi_{\sigma(1)};n_{1,0}\qty(\Phi_{\sigma(2)}))).\label{OCHA2}
\end{align}
If we set $n_{0,1}(L_1)$ to the BRS operator $Q_o(Q_c)$ for open (closed) string field, $n_{1,0}$, $n_{1,1}$ to the Kaku vertices $n^l_{1,0},n^l_{1,1}$, and $n_{2,0}$ to zero map, \eqref{OCHA2} becomes
\begin{align}
    0=n^l_{1,0}\qty(L^\mathrm{lc}_2\qty(\Phi_1,\Phi_2))+(-1)^\abs{\Phi_1}n^l_{1,1}\qty(\Phi_1;n^l_{1,0}\qty(\Phi_2))+(-1)^\abs{\Phi_2}n^l_{1,1}\qty(\Phi_2;n^l_{1,0}\qty(\Phi_1)).\label{OCHA2red}
\end{align}
We can show that the multilinear maps $n^l_{1,0},L^\mathrm{lc}_2$ and $n^l_{1,1}$ satisfy this equation. The diagrams corresponding to each term are shown in Figure \ref{figure:OCHA}.
\begin{figure}
    \begin{minipage}[b]{0.3\linewidth}
        \begin{tikzpicture}[scale=2.25]
            \coordinate (o) at (0,0);
            \coordinate (i) at ($(o)+(pi*2/5 r:0.3)+(0.5,0)$);
            \coordinate (i2) at ($(i)+(0,{0.7*sin(pi*2/5 r)})$);
            \coordinate (v) at ($(i)+(-pi*3/5 r:0.3)$);
            \coordinate (s1) at ($(o)+(0,0)$);
            \coordinate (p11) at ($(i)+(-0.5,0)$);
            \coordinate (p1c) at ($(p11)+(pi*2/5 r:0.7)$);
            \coordinate (p12) at ($(p1c)+(-pi/2 r:1/3)$);
            \coordinate (s2) at ($(p12)+(-pi/2 r:1/3)$);
            \draw ($(i)!1/2!(i2)+(1/4,0)$) ellipse (1/2*0.7/2 and {0.7*sin(pi*2/5 r)/2});
            \draw[double] (s1)--(v);
            \draw[dotted] (v)--(s2);
            \draw (s1)--(p1c);
            \draw[dotted] (p1c)--(s2);
            \draw (p1c)--(i2)--++(1/4,0);
            \draw (i)--++(1/4,0);
            \draw (v)--(i);
            \coordinate (o) at ($(i)!1/2!(i2)+(0.7,0)$);
            \coordinate (i) at ($(o)+(0.5*0.7/2,1*0.7/2)$);
            \coordinate (i2) at ($(i)+(0,-2*0.7/2)$);
            \coordinate (c1) at (o);
            \draw ($(c1)+(0,1*0.7/2)$) arc [start  angle=90,end  angle=270,x  radius=1/2*0.7/2,y  radius=1*0.7/2];
            \draw[dotted] ($(c1)+(0,-1*0.7/2)$) arc [start  angle=-90,end  angle=90,x  radius=1/2*0.7/2,y  radius=1*0.7/2];
            \coordinate (c2) at ($(c1)+(1.25*0.7/2,-1/4*0.7/2)$);
            \draw[name path=C2] (c2) ellipse (1/4*0.7/2 and 3/4*0.7/2);
            \node at ($(c2)+(1/4,-1/3)$) {$\Phi_1$};
            \coordinate (c3) at ($(c1)+(2*0.7/2,1/3*0.7/2)$);
            \draw (c3) ellipse (1/3*0.7/2 and 4/5*0.7/2);
            \node at ($(c3)+(1/4,0)$) {$\Phi_2$};
            \draw ($(c1)+(0,1*0.7/2)$)--(i);
            \draw (i)--($(c2)+(0,3/4*0.7/2)$);
            \draw (i)--($(c3)+(0,4/5*0.7/2)$);
            \draw ($(c1)+(0,-1*0.7/2)$)--(i2);
            \draw (i2)--($(c2)+(0,-3/4*0.7/2)$);
            \draw[white,name path=p] (i2)--($(c3)+(0,-4/5*0.7/2)$);
            \path[name intersections={of= C2 and p}];
            \draw (intersection-2)--($(c3)+(0,-4/5*0.7/2)$);
        \end{tikzpicture}
        \subcaption{$n^l_{1,0}\qty(L^\mathrm{lc}_2\qty(\Phi_1,\Phi_2))$}\label{figure:OCHA1}
    \end{minipage}
    \hspace{1em}
    \begin{minipage}[b]{0.3\linewidth}
        \centering
        \begin{tikzpicture}[scale=1.5]
            \coordinate (o) at (0,0);
            \coordinate (i) at ($(o)+(pi*2/5 r:1)+(0.75,0)$);
            \coordinate (i2) at ($(i)+(0,{sin(pi*2/5 r)})$);
            \coordinate (v) at ($(i)+(-pi*3/5 r:1)$);
            \coordinate (s1) at ($(o)+(0,0)$);
            \coordinate (p11) at ($(s1)+(pi*2/5 r:0.5)$);
            \coordinate (p1c) at ($(p11)+(pi*2/5 r:1.5)$);
            \coordinate (p12) at ($(p1c)+(-pi/2 r:0.7)$);
            \coordinate (s2) at ($(p12)+(-pi/2 r:0.5)$);
            \coordinate (t1) at ($(s2)+(1.5,0)$);
            \coordinate (p21) at ($(t1)+(pi*4/5 r:0.2)$);
            \coordinate (p2c) at ($(i)+(0.75,0)$);
            \coordinate (p22) at ($(p2c)+(-pi*3/5 r:0.5)$);
            \coordinate (t2) at ($(s1)-(s2)+(t1)$);
            \coordinate (c) at ($(i)!1/2!(i2)+(1.5,1/2)$);
            \draw (c) ellipse (1/4 and 1/2);
            \node at ($(c)+(1/2,0)$) {$\Phi_1$};
            \draw[double] (s1)--(t2);
            \draw[dotted] (s2)--(t1);
            \draw (s1)--(p1c);
            \draw[dotted] (p1c)--(s2);
            \draw (t1)--(p2c)--(t2);
            \draw (p1c)--(i2);
            \draw (i)--(p2c);
            \draw (i2)--($(c)+(0,1/2)$);
            \draw (i)--($(c)+(0,-1/2)$);
            %
            \draw[dashed] ($(p11)+(pi*2/5 r:-0.2)$)--($(p22)+(-pi*3/5 r:0.2)$);
            \coordinate (o) at (2,0);
            \coordinate (i) at ($(o)+(pi*2/5 r:0.3)+(0.75,0)$);
            \coordinate (i2) at ($(i)+(0,{0.7*sin(pi*2/5 r)})$);
            \coordinate (v) at ($(i)+(-pi*3/5 r:0.3)$);
            \coordinate (s1) at ($(o)+(0,0)$);
            \coordinate (p11) at ($(i)+(-0.75,0)$);
            \coordinate (p1c) at ($(p11)+(pi*2/5 r:0.7)$);
            \coordinate (p12) at ($(p1c)+(-pi/2 r:1/3)$);
            \coordinate (s2) at ($(p12)+(-pi/2 r:1/3)$);
            \coordinate (c) at ($(i)!1/2!(i2)+(1/2,0)$);
            \draw (c) ellipse (1/2*0.7/2 and {0.7*sin(pi*2/5 r)/2});
            \node at ($(c)+(1/2,0)$) {$\Phi_2$};
            \draw[double] (s1)--(v);
            \draw[dotted] (v)--(s2);
            \draw (s1)--(p1c);
            \draw[dotted] (p1c)--(s2);
            \draw (p1c)--(i2)--++(1/2,0);
            \draw (i)--++(1/2,0);
            \draw (v)--(i);
            %
            \draw[dashed] (p11)--(i);
        \end{tikzpicture}
        \subcaption{$n^l_{1,1}\qty(\Phi_1;n^l_{1,0}\qty(\Phi_2))$}\label{figure:OCHA2}
    \end{minipage}
    \hspace{3em}
    \begin{minipage}[b]{0.3\linewidth}
        \centering
        \begin{tikzpicture}[scale=1.5]
            \coordinate (o) at (0,0);
            \coordinate (i) at ($(o)+(pi*2/5 r:1)+(0.75,0)$);
            \coordinate (i2) at ($(i)+(0,{sin(pi*2/5 r)})$);
            \coordinate (v) at ($(i)+(-pi*3/5 r:1)$);
            \coordinate (s1) at ($(o)+(0,0)$);
            \coordinate (p11) at ($(s1)+(pi*2/5 r:0.5)$);
            \coordinate (p1c) at ($(p11)+(pi*2/5 r:1.5)$);
            \coordinate (p12) at ($(p1c)+(-pi/2 r:0.7)$);
            \coordinate (s2) at ($(p12)+(-pi/2 r:0.5)$);
            \coordinate (t1) at ($(s2)+(1.5,0)$);
            \coordinate (p21) at ($(t1)+(pi*4/5 r:0.2)$);
            \coordinate (p2c) at ($(i)+(0.75,0)$);
            \coordinate (p22) at ($(p2c)+(-pi*3/5 r:0.5)$);
            \coordinate (t2) at ($(s1)-(s2)+(t1)$);
            \coordinate (c) at ($(i)!1/2!(i2)+(1.5,1/2)$);
            \draw (c) ellipse (1/4 and 1/2);
            \node at ($(c)+(1/2,0)$) {$\Phi_2$};
            \draw[double] (s1)--(t2);
            \draw[dotted] (s2)--(t1);
            \draw (s1)--(p1c);
            \draw[dotted] (p1c)--(s2);
            \draw (t1)--(p2c)--(t2);
            \draw (p1c)--(i2);
            \draw (i)--(p2c);
            \draw (i2)--($(c)+(0,1/2)$);
            \draw (i)--($(c)+(0,-1/2)$);
            %
            \draw[dashed] ($(p11)+(pi*2/5 r:-0.2)$)--($(p22)+(-pi*3/5 r:0.2)$);
            \coordinate (o) at (2,0);
            \coordinate (i) at ($(o)+(pi*2/5 r:0.3)+(0.75,0)$);
            \coordinate (i2) at ($(i)+(0,{0.7*sin(pi*2/5 r)})$);
            \coordinate (v) at ($(i)+(-pi*3/5 r:0.3)$);
            \coordinate (s1) at ($(o)+(0,0)$);
            \coordinate (p11) at ($(i)+(-0.75,0)$);
            \coordinate (p1c) at ($(p11)+(pi*2/5 r:0.7)$);
            \coordinate (p12) at ($(p1c)+(-pi/2 r:1/3)$);
            \coordinate (s2) at ($(p12)+(-pi/2 r:1/3)$);
            \coordinate (c) at ($(i)!1/2!(i2)+(1/2,0)$);
            \draw (c) ellipse (1/2*0.7/2 and {0.7*sin(pi*2/5 r)/2});
            \node at ($(c)+(1/2,0)$) {$\Phi_1$};
            \draw[double] (s1)--(v);
            \draw[dotted] (v)--(s2);
            \draw (s1)--(p1c);
            \draw[dotted] (p1c)--(s2);
            \draw (p1c)--(i2)--++(1/2,0);
            \draw (i)--++(1/2,0);
            \draw (v)--(i);
            %
            \draw[dashed] (p11)--(i);
        \end{tikzpicture}
        \subcaption{$n^l_{1,1}\qty(\Phi_2;n^l_{1,0}\qty(\Phi_1))$}\label{figure:OCHA3}
    \end{minipage}
    \caption{The diagrams corresponding to each term in \eqref{OCHA2red} where the case of $\operatorname{sgn}(k_-^1)=\operatorname{sgn}(k_-^2)$.}\label{figure:OCHA}
\end{figure}
There are two moduli for these Riemann surfaces. If we choose Siegel gauge condition as the gauge fixing condition, one of the moduli is twist parameter for the case of Figure \ref{figure:OCHA1}. This parameter measures the amount of relative twist performed before glued between the pairs of closed strings that are being glued. On the other hand, one of the moduli is supplied by the open-open-closed vertex for the cases of Figure \ref{figure:OCHA2} and \ref{figure:OCHA3}. The interaction point of the open-open-closed vertex is placed in the light-cone portion, and as a result, the diagram in Figure \ref{figure:OCHA1} is in  a one-to-one correspondence with the sum of diagrams in Figure \ref{figure:OCHA2} and \ref{figure:OCHA3}. In particular, since we assigned only the momentum $k_-$ as the length of the closed string propagator, the Chan-Paton strips can be seen that this is not significant when we examine whether the OCHA relations hold. In the same way, we consider three cases: $(m,l)=(1,2),(1,3)$ and $(2,1)$. Then, \eqref{OCHA} becomes the followings, respectively.
\begin{align}
    0&=n_{0,1}\qty(n_{1,2}\qty(\Phi_1;\Psi_1,\Psi_2))+n_{1,2}\qty(L_1\qty(\Phi_1);\Psi_1,\Psi_2)+(-1)^\abs{\Phi_1}n_{1,2}\qty(\Phi_1;n_{0,1}\qty(\Psi_1),\Psi_2)\\
    &\qquad\quad+(-1)^{\abs{\Phi_1}+\abs{\Psi_1}}n_{1,2}\qty(\Phi_1;\Psi_1,n_{0,1}\qty(\Psi_2))\\
    &\qquad+n_{0,2}\qty(n_{1,1}\qty(\Phi_1;\Psi_1),\Psi_2)+(-1)^\abs{\Psi_1}n_{0,2}\qty(\Psi_1,n_{1,1}\qty(\Phi_1;\Psi_2))+(-1)^\abs{\Phi_1}n_{1,1}\qty(\Phi_1;n_{0,2}\qty(\Psi_1,\Psi_2))\\
    &\qquad+n_{0,3}\qty(n_{1,0}\qty(\Phi_1),\Psi_1,\Psi_2)+(-1)^\abs{\Psi_1}n_{0,3}\qty(\Psi_1,n_{1,0}\qty(\Phi_1),\Psi_2)+(-1)^{\abs{\Psi_1}+\abs{\Psi_2}}n_{0,3}\qty(\Psi_1,\Psi_2,n_{1,0}\qty(\Phi_1)),\\
    0&=n_{0,1}\qty(n_{1,3}\qty(\Phi_1;\Psi_1,\Psi_2,\Psi_3))+n_{1,3}\qty(L_1\qty(\Phi_1);\Psi_1,\Psi_2,\Psi_3)+(-1)^\abs{\Phi_1}n_{1,3}\qty(\Phi_1;n_{0,1}\qty(\Psi_1),\Psi_2,\Psi_3)\\
    &\qquad\quad+(-1)^{\abs{\Phi_1}+\abs{\Psi_1}}n_{1,3}\qty(\Phi_1;\Psi_1,n_{0,1}\qty(\Psi_2),\Psi_3)+(-1)^{\abs{\Phi_1}+\abs{\Psi_1}+\abs{\Psi_2}}n_{1,3}\qty(\Phi_1;\Psi_1,\Psi_2,n_{0,1}\qty(\Psi_3))\\
    &\qquad+n_{0,2}\qty(n_{1,2}\qty(\Phi_1;\Psi_1,\Psi_2),\Psi_3)+(-1)^\abs{\Psi_1}n_{0,2}\qty(\Psi_1,n_{1,2}\qty(\Phi_1;\Psi_2,\Psi_3))\\
    &\qquad\quad+(-1)^\abs{\Phi_1}n_{1,2}\qty(\Phi_1;n_{0,2}\qty(\Psi_1,\Psi_2),\Psi_3)+(-1)^{\abs{\Phi_1}+\abs{\Psi_1}}n_{1,2}\qty(\Phi_1;\Psi_1,n_{0,2}\qty(\Psi_2,\Psi_3))\\
    &\qquad+n_{0,3}\qty(n_{1,1}\qty(\Phi_1;\Psi_1),\Psi_2,\Psi_3)+(-1)^\abs{\Psi_1}n_{0,3}\qty(\Psi_1,n_{1,1}\qty(\Phi_1;\Psi_2),\Psi_3)+(-1)^{\abs{\Psi_1}+\abs{\Psi_2}}n_{0,3}\qty(\Psi_1,\Psi_2,n_{1,1}\qty(\Phi_1;\Psi_3))\\
    &\qquad\quad+(-1)^\abs{\Phi_1}n_{1,4}\qty(\Phi_1;n_{0,3}\qty(\Psi_1,\Psi_2,\Psi_3))\\
    &\qquad+n_{0,4}\qty(n_{1,0}\qty(\Phi_1),\Psi_1,\Psi_2,\Psi_3)+(-1)^\abs{\Psi_1}n_{0,4}\qty(\Psi_1,n_{1,0}\qty(\Phi_1),\Psi_2,\Psi_3)\\
    &\qquad\quad+(-1)^{\abs{\Psi_1}+\abs{\Psi_2}}n_{0,4}\qty(\Psi_1,\Psi_2,n_{1,0}\qty(\Phi_1),\Psi_3)+(-1)^{\abs{\Psi_1}+\abs{\Psi_2}+\abs{\Psi_3}}n_{0,4}\qty(\Psi_1,\Psi_2,\Psi_3,n_{1,0}\qty(\Phi_1)),\\
    0&=\sum_\sigma(-1)^{\epsilon(\sigma)}\left(\frac{1}{2}n_{0,1}\qty(n_{2,1}\qty(\Phi_{\sigma(1)},\Phi_{\sigma(2)};\Psi_1))+n_{2,1}\qty(L_1\qty(\Phi_{\sigma(1)}),\Phi_{\sigma(2)};\Psi_1)\right.\\
    &\qquad\quad\left.+(-1)^{\abs{\Phi_{\sigma(1)}}+\abs{\Phi_{\sigma(2)}}}\frac{1}{2}n_{2,1}\qty(\Phi_{\sigma(1)},\Phi_{\sigma(2)};n_{0,1}\qty(\Psi_1))\right)\\
    &\qquad+\sum_\sigma(-1)^{\epsilon(\sigma)}\frac{1}{2}\qty(n_{0,2}\qty(n_{2,0}\qty(\Phi_{\sigma(1)},\Phi_{\sigma(2)}),\Psi_1)+(-1)^{\abs{\Psi_1}\qty(1+\abs{\Phi_{\sigma(1)}}+\abs{\Phi_{\sigma(2)}})}n_{0,2}\qty(\Psi_1,n_{2,0}\qty(\Phi_{\sigma(1)},\Phi_{\sigma(2)})))\\
    &\qquad+\sum_\sigma(-1)^{\epsilon(\sigma)}\qty(\frac{1}{2}n_{1,1}\qty(L_2\qty(\Phi_{\sigma(1)},\Phi_{\sigma(2)});\Psi_1)+(-1)^\abs{\Phi_{\sigma(1)}}n_{1,1}\qty(\Phi_{\sigma(1)};n_{1,1}\qty(\Phi_{\sigma(2)};\Psi_1)))\\
    &\qquad+\sum_\sigma(-1)^{\epsilon(\sigma)+\abs{\Phi_{\sigma(1)}}}\qty(n_{1,2}\qty(\Phi_{\sigma(1)};n_{1,0}\qty(\Phi_{\sigma(2)}),\Psi_1)+(-1)^{\abs{\Psi_1}\qty(1+\abs{\Phi_{\sigma(2)}})}n_{1,2}\qty(\Phi_{\sigma(1)};\Psi_1,n_{1,0}\qty(\Phi_{\sigma(2)}))).
\end{align}
If we set the multiliner maps to the Kaku vertices, the light-cone vertex and zero maps, they become
\begin{align}
    0&=n^l_{0,2}\qty(n^l_{1,1}\qty(\Phi_1;\Psi_1),\Psi_2)+(-1)^\abs{\Psi_1}n^l_{0,2}\qty(\Psi_1,n^l_{1,1}\qty(\Phi_1;\Psi_2))+(-1)^\abs{\Phi_1}n^l_{1,1}\qty(\Phi_1;n^l_{0,2}\qty(\Psi_1,\Psi_2))\\
    &\qquad+n^l_{0,3}\qty(n^l_{1,0}\qty(\Phi_1),\Psi_1,\Psi_2)+(-1)^\abs{\Psi_1}n^l_{0,3}\qty(\Psi_1,n^l_{1,0}\qty(\Phi_1),\Psi_2)+(-1)^{\abs{\Psi_1}+\abs{\Psi_2}}n^l_{0,3}\qty(\Psi_1,\Psi_2,n^l_{1,0}\qty(\Phi_1)),\\
    0&=n^l_{0,3}\qty(n^l_{1,1}\qty(\Phi_1;\Psi_1),\Psi_2,\Psi_3)+(-1)^\abs{\Psi_1}n^l_{0,3}\qty(\Psi_1,n^l_{1,1}\qty(\Phi_1;\Psi_2),\Psi_3)+(-1)^{\abs{\Psi_1}+\abs{\Psi_2}}n^l_{0,3}\qty(\Psi_1,\Psi_2,n^l_{1,1}\qty(\Phi_1;\Psi_3)),\\
    0&=n^l_{1,1}\qty(L^\mathrm{lc}_2\qty(\Phi_1,\Phi_2);\Psi_1)+(-1)^\abs{\Phi_1}n^l_{1,1}\qty(\Phi_1;n^l_{1,1}\qty(\Phi_2;\Psi_1))+(-1)^{\qty(\abs{\Phi_1}+1)\abs{\Phi_2}}n^l_{1,1}\qty(\Phi_2;n^l_{1,1}\qty(\Phi_1;\Psi_1)).
\end{align}
In the same way as we did for the case $(m,l)=(2,0)$, we can show that the above relations hold. Additionally, other relations hold trivially. For example, by setting $m=2,l=2$, \eqref{OCHA} becomes
\begin{align}
    0&=\sum_\sigma(-1)^{\epsilon(\sigma)}\left(\frac{1}{2}n_{0,1}\qty(n_{2,2}\qty(\Phi_{\sigma(1)},\Phi_{\sigma(2)};\Psi_1,\Psi_2))+n_{2,2}\qty(L_1\qty(\Phi_{\sigma(1)}),\Phi_{\sigma(2)};\Psi_1,\Psi_2)\right.\\
    &\qquad\quad+(-1)^{\abs{\Phi_{\sigma(1)}}+\abs{\Phi_{\sigma(2)}}}\frac{1}{2}n_{2,2}\qty(\Phi_{\sigma(1)},\Phi_{\sigma(2)};n_{0,1}\qty(\Psi_1),\Psi_2)+\\
    &\qquad\quad\left.+(-1)^{\abs{\Phi_{\sigma(1)}}+\abs{\Phi_{\sigma(2)}}+\abs{\Psi_1}}\frac{1}{2}n_{2,2}\qty(\Phi_{\sigma(1)},\Phi_{\sigma(2)};\Psi_1,n_{0,1}\qty(\Psi_2))\right)\\
    &\qquad+\sum_\sigma(-1)^{\epsilon(\sigma)}\frac{1}{2}\qty(n_{0,2}\qty(n_{2,1}\qty(\Phi_{\sigma(1)},\Phi_{\sigma(2)};\Psi_1),\Psi_2)+(-1)^{\abs{\Psi_1}\qty(1+\abs{\Phi_{\sigma(1)}}+\abs{\Phi_{\sigma(2)}})}n_{0,2}\qty(\Psi_1,n_{2,1}\qty(\Phi_{\sigma(1)},\Phi_{\sigma(2)};\Psi_2)))\\
    &\qquad+\sum_\sigma(-1)^{\epsilon(\sigma)}\left(\frac{1}{2}n_{0,3}\qty(n_{2,0}\qty(\Phi_{\sigma(1)},\Phi_{\sigma(2)}),\Psi_1,\Psi_2)\right.\\
    &\qquad\quad+(-1)^{\abs{\Psi_1}\qty(1+\abs{\Phi_{\sigma(1)}}+\abs{\Phi_{\sigma(2)}})}\frac{1}{2}n_{0,3}\qty(\Psi_1,n_{2,0}\qty(\Phi_{\sigma(1)},\Phi_{\sigma(2)}),\Psi_2)\\
    &\qquad\quad\left.+(-1)^{\qty(\abs{\Psi_1}+\abs{\Psi_2})\qty(1+\abs{\Phi_{\sigma(1)}}+\abs{\Phi_{\sigma(2)}})}\frac{1}{2}n_{0,3}\qty(\Psi_1,\Psi_2,n_{2,0}\qty(\Phi_{\sigma(1)},\Phi_{\sigma(2)}))\right)\\
    &\qquad+\sum_\sigma(-1)^{\epsilon(\sigma)}\left((-1)^\abs{\Phi_{\sigma(1)}}n_{1,1}\qty(\Phi_{\sigma(1)};n_{1,2}\qty(\Phi_{\sigma(2)};\Psi_1,\Psi_2))+\frac{1}{2}n_{1,2}\qty(L_2\qty(\Phi_{\sigma(1)},\Phi_{\sigma(2)});\Psi_1,\Psi_2)\right.\\
    &\qquad\quad\left.+(-1)^\abs{\Phi_{\sigma(1)}}n_{1,2}\qty(\Phi_{\sigma(1)};n_{1,1}\qty(\Phi_{\sigma(2)};\Psi_1),\Psi_2)+(-1)^{\abs{\Phi_{\sigma(1)}}+\abs{\Psi_1}\qty(1+\abs{\Phi_{\sigma(2)}})}n_{1,2}\qty(\Phi_{\sigma(1)};\Psi_1,n_{1,1}\qty(\Phi_{\sigma(2)};\Psi_2))\right)\\
    &\qquad+\sum_\sigma(-1)^{\epsilon(\sigma)}\left((-1)^\abs{\Phi_{\sigma(1)}}n_{1,3}\qty(\Phi_{\sigma(1)};n_{1,0}\qty(\Phi_{\sigma(2)}),\Psi_1,\Psi_2)\right.\\
    &\qquad\quad+(-1)^{\abs{\Phi_{\sigma(1)}}+\abs{\Psi_1}\qty(1+\abs{\Phi_{\sigma(2)}})}n_{1,3}\qty(\Phi_{\sigma(1)};\Psi_1,n_{1,0}\qty(\Phi_{\sigma(2)}),\Psi_2)\\
    &\qquad\quad+(-1)^{\abs{\Phi_{\sigma(1)}}+\qty(\abs{\Psi_1}+\abs{\Psi_2})\qty(1+\abs{\Phi_{\sigma(2)}})}n_{1,3}\qty(\Phi_{\sigma(1)};\Psi_1,\Psi_2,n_{1,0}\qty(\Phi_{\sigma(2)}))\\
    &\qquad\quad\left.+(-1)^{\abs{\Phi_{\sigma(1)}}+\abs{\Phi_{\sigma(2)}}}\frac{1}{2}n_{2,1}\qty(\Phi_{\sigma(1)},\Phi_{\sigma(2)};n_{0,2}\qty(\Psi_1,\Psi_2))\right).
\end{align}
Because multilinear maps $n_{1,2},n_{1,3},n_{2,0},n_{2,1},n_{2,2}$ are set zero maps, it holds trivially. Therefore, other vertices are not necessary, and we have obtained the open-closed Kaku theory action
\begin{align}
    S&=\frac{1}{2}\omega_o(\Psi,Q_o(\Psi))+\frac{1}{3}\omega_o(\Psi,n^l_{0,2}(\Psi,\Psi))+\frac{1}{4}\omega_o(\Psi,n^l_{0,3}(\Psi,\Psi,\Psi))\\
    &\qquad+\omega_o(\Psi,n^l_{1,0}(\Phi))+\frac{1}{2}\omega_o(\Psi,n^l_{1,1}(\Phi;\Psi))\\
    &\qquad+\frac{1}{2}\omega_c(\Phi,Q_c(\Phi))+\frac{1}{3!}\omega_c(\Phi,L^\mathrm{lc}_2(\Phi,\Phi)).
\end{align}

Before proceeding to the next section, we examine the action in the limits $l\to0,\infty$. In the limit $l\to0$, we obtain the action
\begin{align}
    S&=\frac{1}{2}\omega_o(\Psi,Q_o(\Psi))+\frac{1}{3}\omega_o(\Psi,n^\mathrm{lc}_{0,2}(\Psi,\Psi))+\frac{1}{4}\omega_o(\Psi,n^\mathrm{lc}_{0,3}(\Psi,\Psi,\Psi))\\
    &\qquad+\omega_o(\Psi,n^\mathrm{lc}_{1,0}(\Phi))+\frac{1}{2}\omega_o(\Psi,n^\mathrm{lc}_{1,1}(\Phi;\Psi))\\
    &\qquad+\frac{1}{2}\omega_c(\Phi,Q_c(\Phi))+\frac{1}{3!}\omega_c(\Phi,L^\mathrm{lc}_2(\Phi,\Phi)),
\end{align}
where $n^\mathrm{lc}_{p,q}$ is the light-cone vertex. After taking the limit, the theory with the above action has the OCHA structure. Additionally, this coincides with the light-cone open-closed string field theory action except for the kinetic term \cite{Saitoh:1988ew} and with part of unoriented open-closed string field theory action \cite{Kugo:1997rm,Asakawa:1998em}. However there are issues with states where $k_-=0$. As we confirmed in the previous subsection, the open-closed Kaku vertex is not well-defined for the state with $k_-=0$. The open string cubic vertex $n^\mathrm{lc}_{0,2}$, the open-closed vertex $n^\mathrm{lc}_{1,0}$ and the closed string cubic vertex $L^\mathrm{lc}_2$ are not also well-defined for $k_-=0$.

On the other hand, in the limit $l\to\infty$, we obtain the action
\begin{align}
    S&=\frac{1}{2}\omega_o(\Psi,Q_o(\Psi))+\frac{1}{3}\omega_o(\Psi,n^\mathrm{W}_{0,2}(\Psi,\Psi))+\omega_o(\Psi,n^\mathrm{W}_{1,0}(\Phi))\\
    &\qquad+\frac{1}{2}\omega_c(\Phi,Q_c(\Phi))+\frac{1}{3!}\omega_c(\Phi,L^\mathrm{lc}_2(\Phi,\Phi)),
\end{align}
where $n^\mathrm{W}_{0,2}$ is the Witten vertex and $n^\mathrm{W}_{1,0}$ is the open-closed vertex, as given in \cite{Zwiebach:1992bw}. This coincides with the action in \cite{Zwiebach:1992bw} except for the second line. However, since the Witten theory without closed string interactions gives a single cover of moduli space, this action seems somewhat unusual. Furthermore, because the open-closed Kaku vertex is not well-defined for any closed string state in the limit $l\to\infty$, as we confirmed in the previous subsection, we cannot compute closed string amplitudes and so on. Thus, it seems that the open-closed Kaku theory is well-defined only for finite $l$.

\subsection{Field redefinition\label{derive:fieldredefinition}}
In the previous section, we defined the open-closed Kaku theory which depends on the Chan-Paton parameter $l$. In this section, we show that the open-closed Kaku theory with $l+\epsilon$ is equivalent to the open-closed Kaku theory with $l$ following \cite{Erler:2020beb}.

The dynamical string fields of the open-closed Kaku theory with the parameter $l$ will be denoted by $\Psi_l$ and $\Phi_l$. Suppose that there is the following relation between $\Psi_l,\Phi_l$ and $\Psi_{l+\epsilon},\Phi_{l+\epsilon}$,
\begin{align}
    \begin{aligned}
        \Psi_{l+\epsilon}&=\Psi_l-\epsilon\sum_{p,q}\mu^l_{p,q}(\Phi_l^{\wedge p};\Psi_l^{\otimes q}),\\
        \Phi_{l+\epsilon}&=\Phi_l-\epsilon\sum_k\mu^l_k(\Phi_l^{\wedge k}),
    \end{aligned}\label{fieldredefinition}
\end{align}
where $\mu_{p,q}^l$ and $\mu_k^l$ are multilinear maps with cyclicity.
\begin{align}
    \mu_{p,q}^l:\mathcal{H}_c^{\wedge p}\otimes\mathcal{H}_o^{\otimes q}\to\mathcal{H}_o\qc\mu_k^l:\mathcal{H}_c^{\wedge k}\to\mathcal{H}_c
\end{align}
In coalgebra language\footnote{See \cite{Erler:2013xta,Erler:2015uba} for coalgebra.}, they are equivalent to
\begin{align}
    \Psi_{l+\epsilon}&=\Psi_l-\epsilon\pi^o_1\vb*{\mu}^l\qty(e^{\wedge\Phi_l}\otimes\frac{1}{1-\Psi_l}),\\
    \Phi_{l+\epsilon}&=\Phi_l-\epsilon\pi^c_1\vb*{\mu}^l\qty(e^{\wedge\Phi_l}\otimes\frac{1}{1-\Psi_l}),
\end{align}
and
\begin{align}
    \bra{\omega_o}\pi^o_2\vb*{\mu}^l=\bra{\omega_c}\pi^c_2\vb*{\mu}^l=0,
\end{align}
where $\vb*{\mu}^l$ is coderivation formed by the multilinear maps $\mu_{p,q}^l,\mu_k^l$.
\begin{align}
    \vb*{\mu}^l\coloneqq\sum_{p,q\ge0}\vb*{\mu}_{p,q}^l+\sum_{k\ge0}\vb*{\mu}_k^l
\end{align}
Let us also write the multilinear maps $n^l_{p,q}$ 
 and $L^\mathrm{lc}_k$ in coalgebra language.
\begin{align}
    \vb{n}^l&\coloneqq\vb{Q}_o+\vb{n}^l_{0,2}+\vb{n}^l_{0,3}+\vb{n}^l_{1,0}+\vb{n}^l_{1,1}\\
    \vb{L}^\mathrm{lc}&\coloneqq\vb{Q}_c+\vb{L}^\mathrm{lc}_2
\end{align}
If the coderivation $\vb*{\mu}^l$ satisfies 
\begin{align}
    \dv{l}\qty(\vb{n}^l+\vb{L}^\mathrm{lc})=\comm{\vb{n}^l+\vb{L}^\mathrm{lc}}{\vb*{\mu}^l},\label{fieldrelation}
\end{align}
an infinitesimal change of $l$, it gives another string field which is related by the field redefinition \eqref{fieldredefinition}. 

We will expand \eqref{fieldrelation} into component products and solve it. When there are only closed strings in the input and the output of the multilinear maps, \eqref{fieldrelation} implies
\begin{align}
    \dv{l}\vb{L}^\mathrm{lc}=\comm{\vb{L}^\mathrm{lc}}{\sum_{k\ge0}\vb*{\mu}^l_k}.
\end{align}
Because the light-cone vertices do not depend on $l$, we can assume $\vb*{\mu}^l_k=0$. Similarly, when there are only open strings in the input and output, \eqref{fieldrelation} implies
\begin{align}
    0&=\comm{\vb{Q}_o}{\vb*{\mu}^l_{0,1}},\\
    \dv{l}\vb{n}^l_{0,2}&=\comm{\vb{Q}_o}{\vb*{\mu}^l_{0,2}}+\comm{\vb{n}^l_{0,2}}{\vb*{\mu}^l_{0,1}},\\
    \dv{l}\vb{n}^l_{0,3}&=\comm{\vb{Q}_o}{\vb*{\mu}^l_{0,3}}+\comm{\vb{n}^l_{0,2}}{\vb*{\mu}^l_{0,2}}+\comm{\vb{n}^l_{0,3}}{\vb*{\mu}^l_{0,1}},\\
    0&=\comm{\vb{Q}_o}{\vb*{\mu}^l_{0,4}}+\comm{\vb{n}^l_{0,2}}{\vb*{\mu}^l_{0,3}}+\comm{\vb{n}^l_{0,3}}{\vb*{\mu}^l_{0,2}},\\
    &\vdots.
\end{align}
The solution to these equations is already given in \cite{Erler:2020beb}, and we can set
\begin{align}
    \vb*{\mu}^l_{0,q}=0\qfor q\neq2,3,
\end{align}
with $\vb*{\mu}^l_{0,2}$ and $\vb*{\mu}^l_{0,3}$ given by the open string Kaku vertex with appropriate $b$-ghost insertion.

We have to solve the remaining equations.
\begin{align}
    \dv{l}\vb{n}^l_{1,0}&=\vb{Q}_o\vb*{\mu}^l_{1,0}-\vb*{\mu}^l_{1,0}\vb{Q}_c\\
    \dv{l}\vb{n}^l_{1,1}&=\vb{Q}_o\vb*{\mu}^l_{1,1}+\vb{n}^l_{0,2}\vb*{\mu}^l_{1,0}-\vb*{\mu}^l_{0,2}\vb{n}^l_{1,0}-\vb*{\mu}^l_{1,1}\qty(\vb{Q}_o+\vb{Q}_c)\\
    0&=\vb{Q}_o\vb*{\mu}^l_{2,0}+\vb{n}^l_{1,1}\vb*{\mu}^l_{1,0}-\vb*{\mu}^l_{1,1}\vb{n}^l_{1,0}-\vb*{\mu}^l_{1,0}\vb{L}^\mathrm{lc}_2-\vb*{\mu}^l_{2,0}\vb{Q}_c\\
    &\vdots
\end{align}
It is easy to check that we can set
\begin{align}
    \vb*{\mu}^l_{p,q}=0\qfor (p,q)\neq(1,0),(1,1),
\end{align}
and $\vb*{\mu}^l_{1,0},\vb*{\mu}^l_{1,1}$ are given by the open-closed Kaku vertex and the open-open-closed Kaku vertex with appropriate $b$-ghost insertion. We provide a concrete description. We define $e^l_{1,1}$ and $e^l_{1,2}$ as
\begin{align}
    e^l_{1,1}(B_c;A)&=(-1)^{1+(\abs{A}+1)\abs{B_c}}\omega_o(A,\mu^l_{1,0}(B_c)),\\
    e^l_{1,2}(B_c;A,C)&=(-1)^{\abs{A}+\abs{B_c}}\omega_o(A,\mu^l_{1,0}(B_c;C)).
\end{align}
Then the surface states $\bra{e^l_{1,1}}$ and $\bra{e^l_{1,2}}$ are given by
\begin{align}
    \bra{e^l_{1,1}}&=\bra{V^l_{1,1}}b^l_{(1,1),l},\\
    \bra{e^l_{1,2}}&=\int_{m^l_-}^{m^l_+}\bra{\Sigma^{l,m}_{1,2}}b^{l,m}_{(1,2),m}b^{l,m}_{(1,2),l},
\end{align}
where $b^l_{(1,1),l},b^{l,m}_{(1,2),l}$ are contour
integrals of the $b$-ghost:
\begin{align}
    b^l_{(1,1),l}&=b^l_{((1,1),c),l}\otimes\mathbb{I}+\mathbb{I}\otimes b^l_{((1,1),o),l},\\
    b^{l,m}_{(1,2),l}&=b^{l,m}_{((1,2),c),l}\otimes\mathbb{I}^{\otimes2}+\mathbb{I}\otimes b^{l,m}_{((1,2),1),l}\otimes\mathbb{I}+\mathbb{I}^{\otimes2}\otimes b^{l,m}_{((1,2),2),l}.
\end{align}
Using the vector fields
\begin{align}
    v_{((1,1),r),l}^l(u)&=\pdv{(f_{((1,1),r)}^l)^{-1}}{l}\qty(\pdv{(f_{((1,1),r)}^l)^{-1}}{u})^{-1},\\
    v_{((1,2),r),l}^{l,m}(u)&=\pdv{(f_{((1,2),r)}^{l,m})^{-1}}{l}\qty(\pdv{(f_{((1,2),r)}^{l,m})^{-1}}{u})^{-1},
\end{align}
$b^l_{((1,1),r),l}$ and $b^{l,m}_{((1,2),r),l}$ are given by
\begin{align}
    f^l_{((1,1),r)}\circ b^l_{((1,1),r),l}&=\oint_{f^l_{((1,1),r)}(0)}\frac{\dd{u}}{2\pi i}v_{((1,1),r),l)}^l(u)b(u)\\
    f^{l,m}_{((1,2),r)}\circ b^{l,m}_{((1,2),r),l}&=\oint_{f^{l,m}_{((1,2),r)}(0)}\frac{\dd{u}}{2\pi i}v_{((1,2),r),l)}^{l,m}(u)b(u).
\end{align}

Therefore, the dynamical string fields $\Psi_{l+\epsilon}$ and $\Phi_{l+\epsilon}$ of the open-closed Kaku theory with $l+\epsilon$ are related to $\Psi_l$ and $\Phi_l$ by the field redefinition
\begin{align}
    \begin{aligned}
        \Psi_{l+\epsilon}&=\Psi_l-\epsilon\qty(\mu^l_{0,2}(\Psi_l,\Psi_l)+\mu^l_{0,3}(\Psi_l,\Psi_l,\Psi_l)+\mu^l_{1,0}(\Phi_l)+\mu^l_{1,1}(\Phi_l;\Psi_l)),\\
        \Phi_{l+\epsilon}&=\Phi_l,
    \end{aligned}
\end{align}
and the open-closed Kaku theory with $l+\epsilon$ is equivalent to the open-closed Kaku theory with $l$.

The part of the field redefinition $\mu^l_{1,0}$ is given by the open-closed Kaku vertex $n^l_{1,0}$. Since we confirmed in subsection \ref{open-closed} that the open-closed Kaku vertex is not well-defined for the closed string state with $k_-=0$, the field redefinition is not well-defined. This is not surprising, as we have already faced a similar problem in constructing a field redefinition of the open Kaku theory \cite{Erler:2020beb}. Because $\mu^l_{0,2}$ is not well-defined for the state with $k_-=0$ in the limit $l\to0$, it was shown that there is no well-defined mapping for the state with $k_-=0$ in the open Kaku theory with $l=0$. In our case, this also implies that there is no well-defined mapping for the closed string state with $k_-=0$ in the open-closed Kaku theory with any $l$. Particularly, after taking the limit $l\to\infty$, the open-closed Kaku vertex is not well-defined for states with any momentum. Thus, our discussion does not imply that the open-closed Kaku theory with finite $l$ is equivalent to the open-closed Kaku theory with $l\to\infty$.
\section{Closed string amplitude around tachyon vacuum solution\label{amplitude}}
We have already found the tachyon vacuum solution \cite{Ando:2023cnx}. Furthermore, we obtained the open-closed Kaku theory in the previous section. In this section, we consider the open-closed Kaku theory around the tachyon vacuum solution. Its action is given by
\begin{align}
    S^l_\text{tv}&=\frac{1}{2}\omega_o(\Psi,Q_o^{l,\text{tv}}\Psi)+\frac{1}{3}\omega_o(\Psi,n_{0,2}^l(\Psi,\Psi))+\frac{1}{4}\omega_o(\Psi,n_{0,3}^l(\Psi,\Psi,\Psi))\\
    &\qquad+\omega_o(\Psi,n_{1,0}^l(\Phi))+\frac{1}{2}\omega_o(\Psi,n_{1,1}^l(\Phi;\Psi)\\
    &\qquad+\frac{1}{2}\omega_c(\Phi,Q_c\Phi)+\frac{1}{3!}\omega_c(\Phi,L_2^\mathrm{lc}(\Phi,\Phi)),
\end{align}
where $\Psi_\text{tv}$ is the tachyon vacuum solution and $Q_o^{l,\text{tv}}$ is defined by
\begin{align}
    Q_o^{l,\text{tv}}A\coloneqq Q_oA+n_{0,2}^l(\Psi_\text{tv},A)+n_{0,2}^l(A,\Psi_\text{tv}).
\end{align}
We will compute closed string amplitudes around the tachyon vacuum solution. To do this, it is necessary to choose a gauge condition and a propagator. Here we do not specify a particular gauge condition and we write $h_o$ and $h_c$ as propagators of the open string field and the closed string field in a given gauge condition.

First, we consider the tree-level 3-point amplitude around the solution. The action includes the closed string cubic vertex, the open-closed vertex and the open-open-closed vertex. Hence, the amplitude is given by
\begin{align}
    \mathcal{A}_{3-\text{closed}}&=\omega_c(\Phi,L_2^\mathrm{lc}(\Phi,\Phi))+2\omega_o\qty(n_{1,0}^l(\Phi),h_on_{0,2}^l(h_on_{1,0}^l(\Phi),h_on_{1,0}^l(\Phi)))+3\omega_o\qty(n_{1,0}^l(\Phi),h_on_{1,1}^l(\Phi;h_on_{1,0}^l(\Phi))).
\end{align}
Of course, the external states are on-shell
\begin{align}
    Q_c\Phi=0.
\end{align}
Because of one of the OCHA relations \eqref{OCHA}
\begin{align}
    Q^{l,\mathrm{tv}}_o(n_{1,0}^l(\Phi))+n_{1,0}^l(Q_c(\Phi))=0,\label{relation:shiftOCHA}
\end{align}
we obtain
\begin{align}
    Q^{l,\mathrm{tv}}_o(n_{1,0}^l(\Phi))=0.
\end{align}
Since the cohomology of $Q^{l,\mathrm{tv}}_o$ vanishes, $n_{1,0}(\Phi)$ is a $Q^{l,\mathrm{tv}}_o$-exact state. As a result, the tree-level 3-point amplitude around the tachyon vacuum solution is given by
\begin{align}
    \mathcal{A}_{3-\text{closed}}=\omega_c(\Phi,L_2^\mathrm{lc}(\Phi,\Phi)).
\end{align}
As expected, this is the tree-level 3-point pure closed string amplitude.

Finally, we compute the tree-level $n$-point amplitude around the solution. The amplitudes consist of three types. One of them is amplitudes constructed solely from closed string vertices. For example, this includes
\begin{align}
    \omega_c(\Phi,L_2^\mathrm{lc}(\Phi,\Phi))\qc\omega_c(\Phi,L_2^\mathrm{lc}(L_2^\mathrm{lc}(\Phi,\Phi),h_c\Phi)).
\end{align}
Another type consists of amplitudes that include the open-closed vertex e.g.
\begin{align}
    \omega_o\qty(n_{1,0}^l(\Phi),h_on_{0,2}^l(h_on_{1,0}^l(\Phi),h_on_{1,0}^l(\Phi)))\qc\omega_o\qty(n_{1,0}^l(\Phi),h_on_{1,1}^l(\Phi;h_on_{1,0}^l(\Phi))).
\end{align}
These do not contribute to the amplitudes due to the above discussion. The last type includes amplitudes that do not involve the open-closed vertex. However, this does not contribute to the tree-level amplitudes. It is easy to show that the Feynman diagrams corresponding to this type contain at least one loop. Therefore, the tree-level $n$-point amplitude around the tachyon vacuum solution is the tree-level $n$-point pure closed string amplitude.
\section{Summary\label{Summary}}
In this paper, we defined the open-closed Kaku vertex and the open-open-closed Kaku vertex. These vertices depend on the Chan-Paton parameter $l$ similar to the open string Kaku vertex, and they coincide with the light-cone vertex \cite{Saitoh:1988ew,Kugo:1997rm,Asakawa:1998em} and the vertex \cite{Shapiro:1987ac,Zwiebach:1992bw} in the limits $l\to0$ and $l\to\infty$, respectively. Using the Kaku vertices, we obtained the open-closed Kaku theory action. This theory has the open-closed homotopy algebra structure, and an infinitesimal change in $l$ leads to another open-closed string field theory, just like the open string Kaku theory. However we must be careful in the limit $l\to\infty$. After taking this limit, the open-closed vertex is not well-defined, and we cannot compute closed string amplitudes and so on.

The most important aspect is that the Kaku theory is a framework in which the tachyon vacuum solution can be found and open-closed interactions can be treated. Hence, we confirmed that closed string amplitudes around the tachyon vacuum solution are pure closed string amplitudes as expected.

An interesting point is that we obtained the expected amplitudes by using only one of the OCHA relation \eqref{relation:shiftOCHA} and that there is no open string excitations around the tachyon vacuum solution in the open-closed Kaku theory. However, in general theory, it is not clear. For example, if an action includes open-closed-closed vertex $n_{2,0}$, the following would contribute the four-point tree amplitude.
\begin{align}
    \omega_o(n_{2,0}(\Phi,\Phi),h_on_{2,0}(\Phi,\Phi))
\end{align}
We have shown only that $n_{1,0}(\Phi)$ is $Q^\mathrm{tv}$-exact. We do not know whether $n_{2,0}(\Phi,\Phi)$ is $Q^\mathrm{tv}$-exact, but this should not contribute around the tachyon vacuum solution. Using the OCHA relation, even if an action includes higher interaction vertices, we may still achieve the same result.

In the open-closed Kaku theory, there are many open problems. One of them is the relation between the open-closed Kaku theory and the Witten theory. We still do not know whether the open-closed Kaku theory is equivalent to the Witten theory because the field redefinition is not well-defined. If we understand the relation, it may provide useful insights into computation of closed string amplitudes around the tachyon vacuum solution in Witten theory. To do this, we should carefully observe what happens in the limit $l\to\infty$, and it may be necessary to modify the Witten theory.
\section*{Acknowledgments}
The author would like to thank Nobuyuki Ishibashi for reading a preliminary draft and helpful comments.
This work was supported by JST, the establishment of university fellowships towards the creation of science technology innovation, Grant Number JPMJFS2106.
\printbibliography
\end{document}